\documentclass[prl,twocolumn,showpacs]{revtex4}
\usepackage{graphics}
\usepackage{color}
\usepackage{amsmath,amssymb,amsthm}
\usepackage{times}
\definecolor{wine}{rgb}{0.5,0.1,0.1}
\begin{document}

%\title{lattice Boltzmann model of vesicle deforming and  moving in fluid}
\title{Shape changes and motion of a vesicle in a fluid using a lattice
    Boltzmann model}
\author{Huabing Li$^{1,2}$, Houhui Yi$^{1,3}$, Xiaowen Shan$^{4}$, and
  Haiping Fang$^{1}$\footnote{To whom correspondence should be
  addressed. Email address: fanghaiping@sinap.ac.cn}
}

\affiliation{
  $^1$Shanghai Institute of Applied Physics, Chinese Academy of
  Sciences, P.O. Box 800-204, Shanghai 201800, China\\
  $^2$Department of information material science and engineering,
  Guilin University of Electronic Technology, Guilin 541004, China\\
  $^3$Graduate School of the Chinese Academy of Sciences, Beijing
  100080, China \\
  $^4$EXA Corporation, 3 Burlington Woods Drive, Burlington, MA 01803, USA
}
%\date{}F}

\begin{abstract}
We study the deformation and motion of an erythrocyte in fluid
flows via a lattice Boltzmann method. To this purpose, the bending
rigidity and the elastic modulus of isotropic dilation  are
introduced and incorporated with the lattice Boltzmann simulation,
and the membrane-flow interactions on both sides of the membrane
are carefully examined. We find that the static biconcave shape of
an erythrocyte is quite stable and can effectively resist the
pathological changes on their membrane. Further, our simulation
results show that in shear flow, the erythrocyte will be highly
flattened and undergo tank tread-like motion. This phenomenon has
been observed by experiment very long time ago, but it has feazed
the boundary integral and singularity methods up to the present.
Because of its intrinsically parallel dynamics, this lattice
Boltzmann method is expected to find wide applications for both
single and multi-vesicles suspension as well as complex open
membranes in various fluid flows for a wide range of Reynolds
numbers.
\end{abstract}
\maketitle
 Vesicle whose membrane consists of lipid bilayer is essential to the function of
biological systems \cite{Fung97}. Erythrocyte is the most
important kind of vesicles. In the past 30 years, the dynamics of
vesicle has received particular
attention~\cite{Ou-Yang87,Lipowsky96,PozrikidisBook92,Secomb01}.
It has been recognized that the equilibrium shape can be obtained
by minimizing the bending energy of the membrane \cite{Ou-Yang87}.
However, the studies on the unsteady states lag behind, mainly
because of the numerical difficulty on capturing the coupling
between the membrane and ambient fluids while the vesicle is
deforming and moving under the hydrodynamic forces exerted on both
sides of its elastic membrane \cite{Lipowsky96}.
%Molecular dynamic method (MD) has been applied to this problem but
%without any real fluid flow~\cite{Moldovan99}.
When the vesicle is very close to a solid static boundary and the
Reynolds number is very small, lubrication theory has been
extended to the study \cite{Secomb01}. Recently, the deformations
of vesicles in the approximation of Stokes flow has been
extensively studied by the boundary integral and singularity
methods \cite{PozrikidisBook92,Lipowsky96}, which has a limitation
imputed to the absence of numerical smoothing and accuracy of
regridding \cite{Pozrikidis03} so that it can not describe the
highly flattened tank-treading shapes.

Erythrocytes in large blood vessels, which have larger Reynolds
numbers, not only affect the viscosity of the
fluid~\cite{Stoltz99}, but also often subject to pathological
changes of their membranes due to the large shear stress
\cite{Gallucci99}. This calls for computing models for
deformations of vesicles, particularly with inhomogeneous
membrane, in fluid flow with a wide range of Reynolds numbers.
Moreover, considering the numerical complexity of the vesicle
deformations and multi-vesicle suspensions, and the recent
development of computational technique, especially on the PC
clusters and internet grid computing, it is most desirable that
the numerical models are intrinsically parallel. The lattice
Boltzmann method (LBM)~\cite{McNamara88,Qian92} has localized
kinetic nature which is not only intrinsically parallel but also
easy to capture the interaction between a fluid and a small
segment of a deformable boundary. In the past fifteen years, the
LBM has been recognized as an alternate method for computational
fluid dynamics, especially in the areas of complex fluids such as
particle suspension flow~\cite{Ladd01}, binary
mixture~\cite{Gunst91} and blood flow~\cite{Buick02}. However,
applicability of LBM on the vesicle deforming and moving remains
an open problem. A crucial difficulty is the physics that the
membrane is viscoelstic and area-unchangeable while the total
volume of the vesicle keeps a constant. Numerically, the
viscoelstic property demands that the boundary condition and the
hydrodynamic force should be accurately treated even in a very
small segment. In two-dimensional problems, area- and volume-
unchangeability are reduced to inextensibility of the
one-dimensional membrane and area-unchangablility of the area
surrounded by the one-dimensional membrane. In this Letter we
develop a lattice Boltzmann model in which the elastic properties,
length constrain of the membrane and volume constrain on area of
vesicle incorporated in a discrete way with a boundary treatment
for moving boundaries.

We choose to work with the D2Q9 model on a two-dimensional square
lattice with nine velocities~\cite{Qian92}. Let $f_\alpha({\bf
x},t)$ be the non-negative distribution function which can be
thought as the number of fluid particles at site ${\bf x}$, time
$t$, and possessing one of the nine velocities ${\bf e}_\alpha$.
The distribution function evolves according to a Boltzmann
equation that is discrete in both space and time,
\begin{equation}
  f_\alpha({\bf x}+{\bf e}_\alpha,t+1)-f_\alpha({\bf x},t) = -\frac 1\tau(f_\alpha-f_\alpha^{eq}).
  \label{1}
\end{equation}
The density $\rho$ and macroscopic velocity ${\bf u}$ are defined
by
\begin{equation}
  \label{rho}
  \rho = \sum_\alpha f_\alpha, \ \ \ \rho {\bf u} = \sum_\alpha f_\alpha{\bf e}_\alpha.
\end{equation}
Here, the equilibrium distribution function $f_\alpha^{eq}$
depends only on the local density $\rho$ and flow velocity ${\bf
u}$. The pressure and the viscosity are defined by the equations
$p = c_s^2\rho$ with $c_s^2 = 1/3$ and $\nu = {(2\tau-1)/6}$,
respectively~\cite{Qian92}.

The membrane of the erythrocytes has bending rigidity potential,
which can be written as~\cite{Ou-Yang87,PozrikidisBook92}
\begin{equation}
  \phi_B = \frac{1}{2}k_B\int{k^2 \mathrm{d}l},\label{Energy}
\end{equation}
where $k_{B}$ is the bending modulus, $k$ and $l$ are the
curvature and the arc length of the membrane, separately.
Biomembranes are formed by a lipid bilayer, which is viscoelastic.
The erythrocyte viscoelasticity is usually assumed to be
Kelvin-Voigt \cite{Evans76} and described by
\begin{equation}
    T_{mn} = 2\eta\dot{\varepsilon}_{mn},\label{Visco}
\end{equation}
where $T_{mn}$ is the viscous stress, $\dot{\varepsilon}_{mn}$ and
$\eta$ are the strain rate and the viscous coefficient of the
membrane. The viscoelasticity of the membrane does not change the
static shapes of erythrocytes.

Numerically, the membrane of a two-dimensional erythrocyte is
discretized into equilength segments.  We implement a no-slip
boundary condition and compute the hydrodynamic forces on both
sides of each segment according to the scheme proposed recently
\cite{Li2004b}. The isotropic dilation can be described by
adopting an in-plane potential $\phi_k$ between neighboring
segments as
\begin{equation}
  \phi_k = \frac{1}{2}{k_{k}\sum_{i = 1}^{N}(l_i-l_{0})^2},
\end{equation}
where $k_{k}$ is the elastic coefficient of the membrane, $l_{0}$
and $l_{i}$ are the original and simulated length of segment $i$
respectively. Pozrikidis has already shown that the transverse
shear tension due to the bending energy can be computed as
\cite{PozrikidisBook92}
\begin{equation}
  F^B = k_B\frac{\partial k}{\partial l}.
\end{equation}
The force exerted on segment $i$ on the normal direction due to
the bending energy is
\begin{equation}
  F_{i}^B = k_B\frac{k_{i+1}-k_i}{l_i},
\end{equation}
where $k_{i+1}$ and $k_i$ are the curvatures of the membrane at
segments $i+1$ and $i$, respectively. The membrane viscous
resistance on the normal direction on segment $i$ is
\begin{equation}
    F^r_i = -\eta (v_{i+1, n}-v_{i, n}),
\end{equation}
where $v_{i+1, n}$ and $v_{i, n}$ is the velocity of segments
$i+1$ and $i$ along the normal direction of segment $i$,
separately.

The translation of each segment is updated at each Newtonian
dynamics time step according to the sum of all the forces on the
segment by using a so-called half-step `leap-frog'
scheme~\cite{Allen1987}.

The membrane parameters are set to be $k_{B} = 1.8 \times
10^{-12}$ $\mathrm{dyn\cdot cm}$~\cite{PozrikidisBook92} and $\eta
= 1.0\times 10^{-5}$ $\mathrm{dyn\cdot s/cm}$~\cite{Evans76}. The
blood serum is usually assumed to be Newtonian and has a viscosity
$\nu = 0.01$ $\mathrm{cm^2/s}$ and density $\rho = 1.00$
$\mathrm{g/cm^3}$~\cite{PozrikidisBook92}. The fluid inside the
erythrocytes is also assumed to be serum too. The thickness and
the density of the membrane are set to be 0.02 $\mathrm{\mu m}$
and 1.00 $\mathrm{g/cm^3}$, respectively~\cite{Oka88}. The
cross-membrane pressure drop of an erythrocyte can be expressed by
the chemical potential drop~\cite{Walter95}
\begin{equation}
  \Delta \mu = RT\ln(\frac{p_{\mathrm{out}}}{p_{\mathrm{in}}}),
\end{equation}
where $p_{\mathrm{out}}$ and $p_{\mathrm{in}}$ are the pressure
outside and inside the erythrocyte, separately. The temperature is
set to the human body temperature $37^o$~C. The longer axis of an
erythrocyte is set to 3.9~$\mu$m and the according shorter axis is
0.4~$\mu$ m~\cite{Oka88}. The elastic modulus of isotropic
dilation is $k_{k} = 500$
$\mathrm{dyn/cm}$~\cite{PozrikidisBook92} .
\begin{figure}[htb]
  \begin{center}
    {\scalebox{0.5}[0.5]{\includegraphics*[80,380][490,630]
        {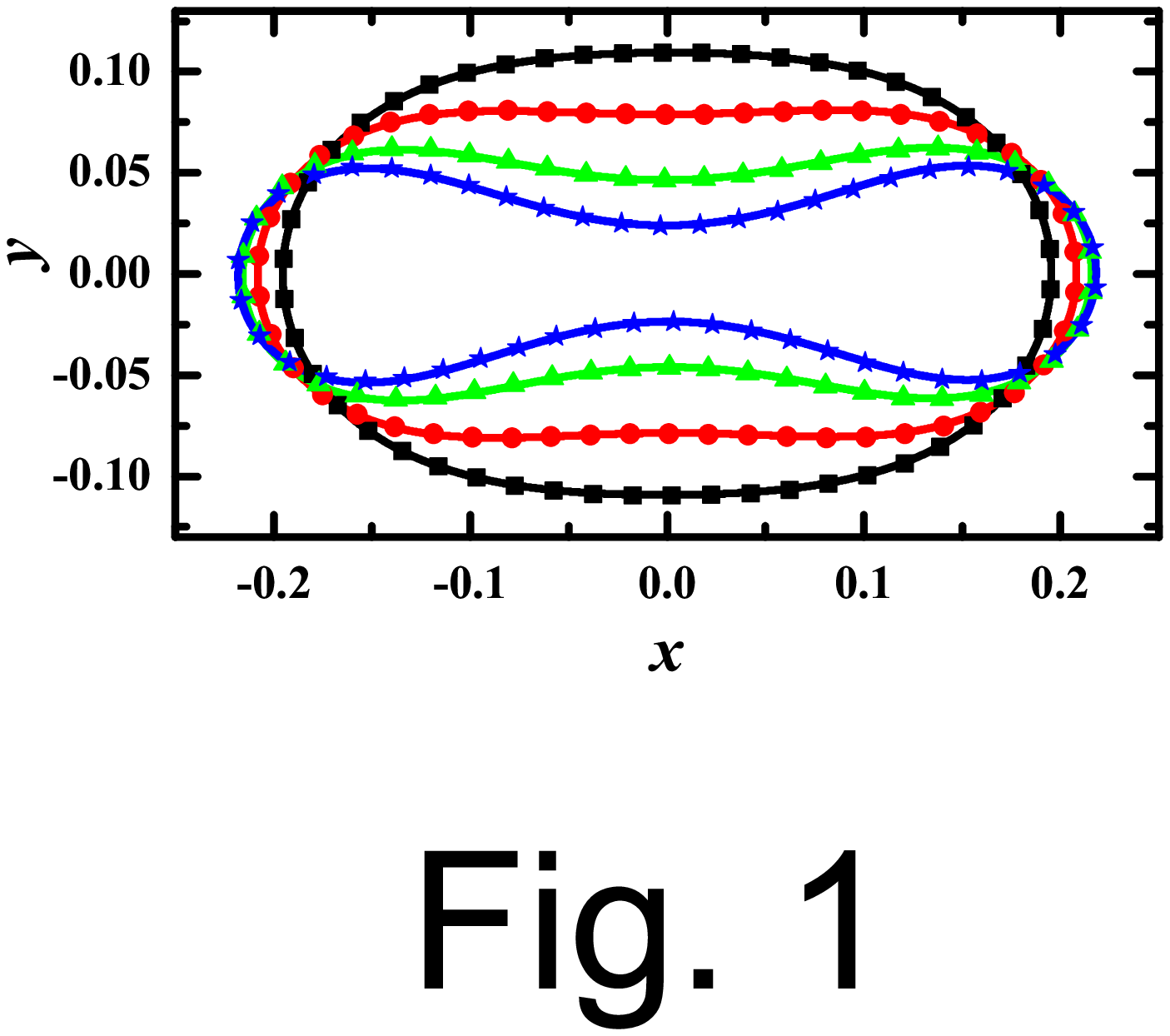}}}
  \end{center}
  \caption{The static profiles of an erythrocyte for $\Delta \mu$ =
  2.71 ($\square$), 2.97 (\textcolor{red}{$\bigcirc$}), 3.39
    ($\textcolor{green}{\star}$), and 3.82 ($\textcolor{blue}{\triangle}$) $\times 10^{-5}$J/mol calculated from
    lattice Boltzmann simulations (symbols) together with those from a shooting
    method~\cite{Pozrikidis02} (lines). $x$ and $y$ are normalized by the
    total length of the membrane. }
  \label{shape}
\end{figure}

The simulation domain consists of $200 \times 80$ lattice units.
We have tested that when the width is greater than 200 lattice
units, increase of width dose not affect the simulation results if
the boundary conditions of both left and right sides are periodic.
Initially, the fluid was homogeneous and steady. A circular
membrane placed at the center of the square without stretching was
discretized into $N = 100$ segments. The radius of the initial
circular membrane was 20 lattice units. The length in each lattice
unit corresponded to 0.14~$\mathrm{\mu m}$. The relaxation time
$\tau$ was fixed to be 0.75, resulting in $1.69\times 10^{-9}$~s
for each time step. The initial density of the fluid inside and
outside the close membrane was set to be one lattice Boltzmann
unit. The other non-dimensional quantities relevant to lattice
Boltzmann simulations could be computed correspondingly. In the
simulation, the fluid in the square of $6 \times 6$ lattice units
at the center of the system was pumped out with a speed of
$1.69\times 10^{-9}$ $\mathrm{g/s}$, {\it i.e.}, $1/1000$ per
time-step, until the predetermined chemical potential drop was
reached, the mass inside the erythrocyte is then remained constant
for the remainder of the simulation.

Fig.~\ref{shape} shows the profiles of an erythrocyte with
different chemical potential drops $\Delta \mu$ together with that
of a shooting method~\cite{Pozrikidis02}. As $\Delta \mu$
increases, the erythrocyte deforms from a circle to an ellipse,
and into an biconcave shape. Excellent agreement can be found
between the two methods. In order to further characterize the
agreement, we have also computed the relative global error
$\sigma$ of the curvature of the membrane between the two methods,
defined by
\begin{equation}
  \sigma = {\sum\limits_{i = 1}^{N}(k_i-k'_i)^2}/
         {\sum\limits_{i = 1}^{N}k_i^{'2}},
\end{equation}
where $k_i$ and $k'_i$ are the curvatures at segment $i$,
calculated from lattice Boltzmann simulations and the shooting
method~\cite{Pozrikidis02}, respectively. We found that the
relative global error was less than $10^{-5}$.

\begin{figure}[htb]
  \begin{center}
    {\scalebox{0.35}[0.35]{\includegraphics*[85,155][515,740]
        {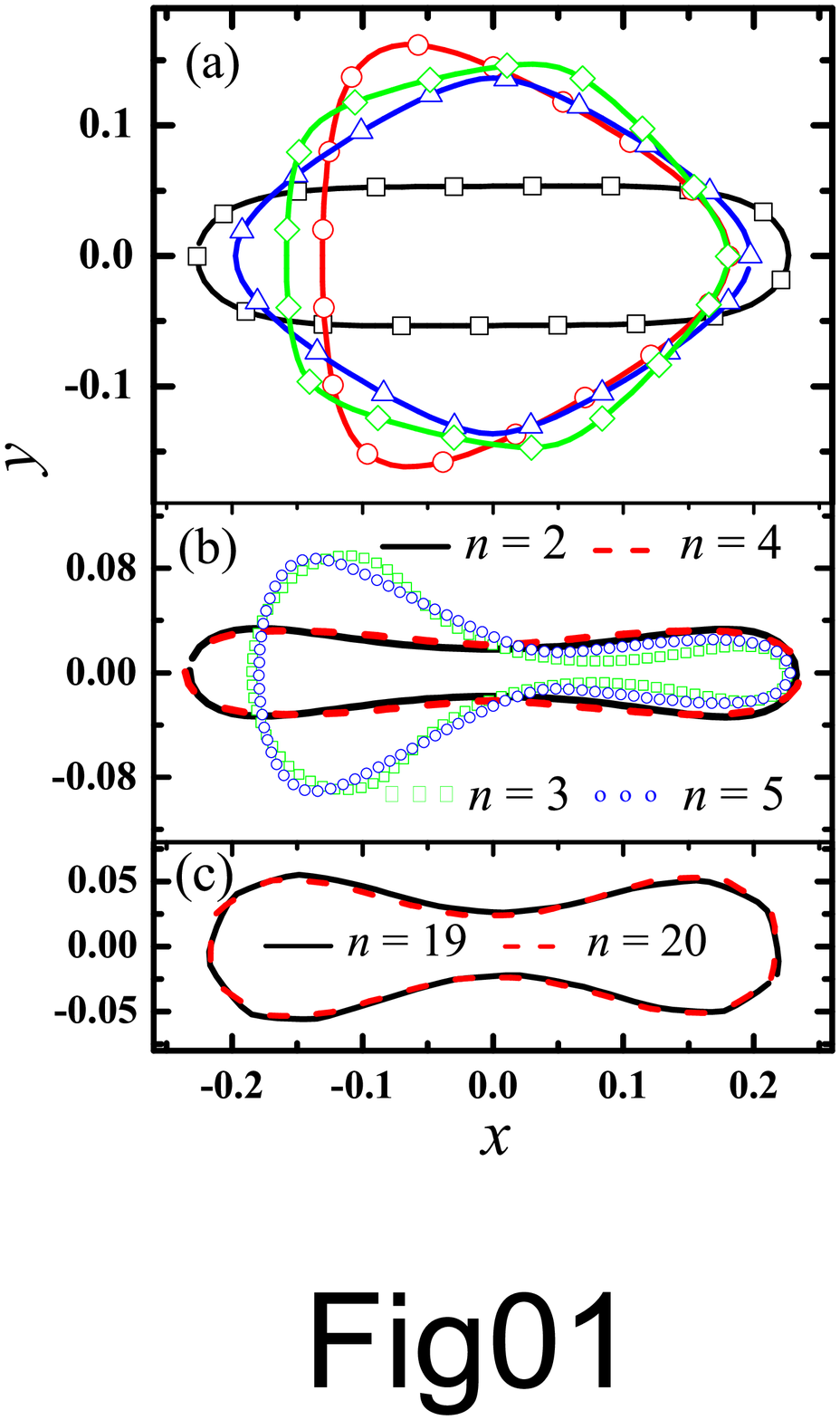}}}
  \end{center}
  \caption{Static profiles of erythrocytes for bending modulus varying
    according to Eq. (\ref{bendingf}). (a) For small chemical potential
    drop $\Delta \mu$ = $4.24\times 10^{-6}$ J/mol, lines with $\square,
    \textcolor{red}{\bigcirc}, \textcolor{blue}{\triangle},
    \textcolor{green}{\diamondsuit}$ correspond to $n = 2, 3, 4, 5$,
    respectively. (b) and (c) Erythrocyte profiles for larger chemical
    potential drops. $\Delta \mu$ = 0.68, 0.68, 1.87, 1.10,
    1.18 and 1.10 ($\times 10^{-5}$J/mol) for $n = 2, 4, 3, 5, 19, 20$.}
  \label{collapse}
\end{figure}

Due to high values of shear stress in the large arteries or
 in cases of oxidant injury \cite{Gallucci99}, erythrocyte membranes can be
pathologically damaged so that the bending rigidity modulus
becomes non-uniform. To study the effect on the shapes of the
erythrocytes, we performed numerical simulations with bending
modulus changing periodically along the membrane
\begin{eqnarray}
  K_B = K_0\left[\delta+(1-\delta)\cos^{2}(n\pi l')\right],
  0 \leq l' <1,
  \label{bendingf}
\end{eqnarray}
where $\delta$ is a constant, $l'$ is the normalized arc length of
the membrane, and $n$ is an integer. The simulation results for
$\delta = 0.1$ are shown in Fig.~\ref{collapse}. For small
chemical potential drop, say $\Delta \mu \leq 4.24\times 10^{-6}$
J/mol, the shape of the erythrocyte exhibits the same symmetry of
$K_B$. Remarkably, when $\Delta \mu$ is large enough, all
erythrocytes, with different bending modulus, become biconcave
shapes, {\em i.e.}, the biconcave shapes can effectively resist
the perturbations. We note that the perturbations are quite large
as the minimum of $K_B$ is only $10\%$ of its original value.
Further, erythrocyte profiles for perturbation wave numbers of the
same parity are very similar to each other. As shown in
Fig.~\ref{collapse} (b), the difference between the profiles for
$n = 3$ and $n = 5$, or that between $n = 2$ and $n = 4$ is almost
indistinguishable whereas the discrepancy between the cases of $n
= 2$ and $n = 3$ is clear. When the wave number $n$ is large
enough, the difference for odd and even $n$ becomes very small as
shown in Fig.~\ref{collapse} (c). The chemical potential drop
needed to collapse an erythrocyte is smaller for an even $n$ than
that for an odd $n$, this difference becomes vanishingly small as
$n$ increases.
\begin{figure}[htb]
  \begin{center}
    {\scalebox{0.28}[0.28]{\includegraphics*[85,280][482,620]
        {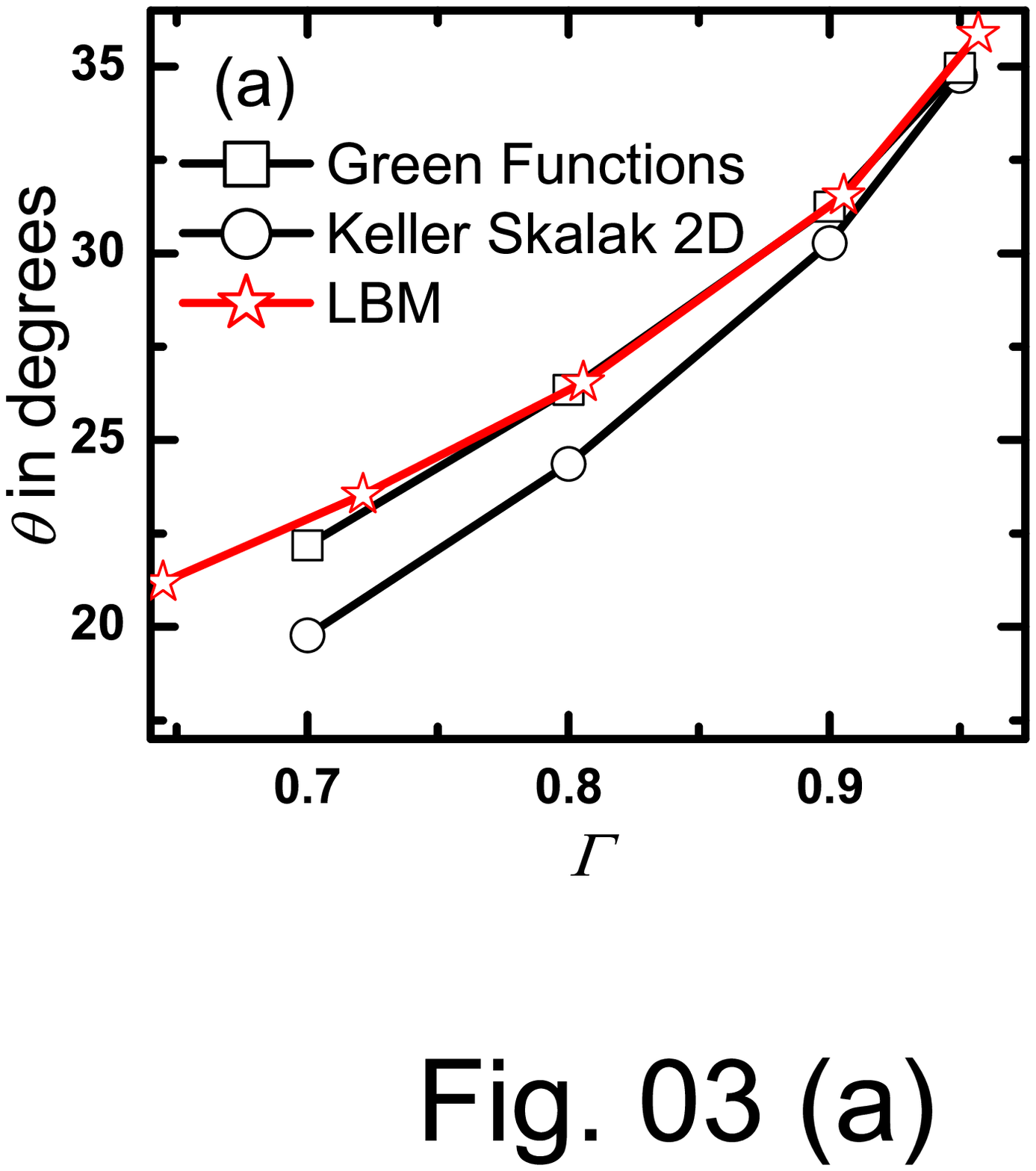}}}
    {\scalebox{0.28}[0.28]{\includegraphics*[80,270][500,630]
        {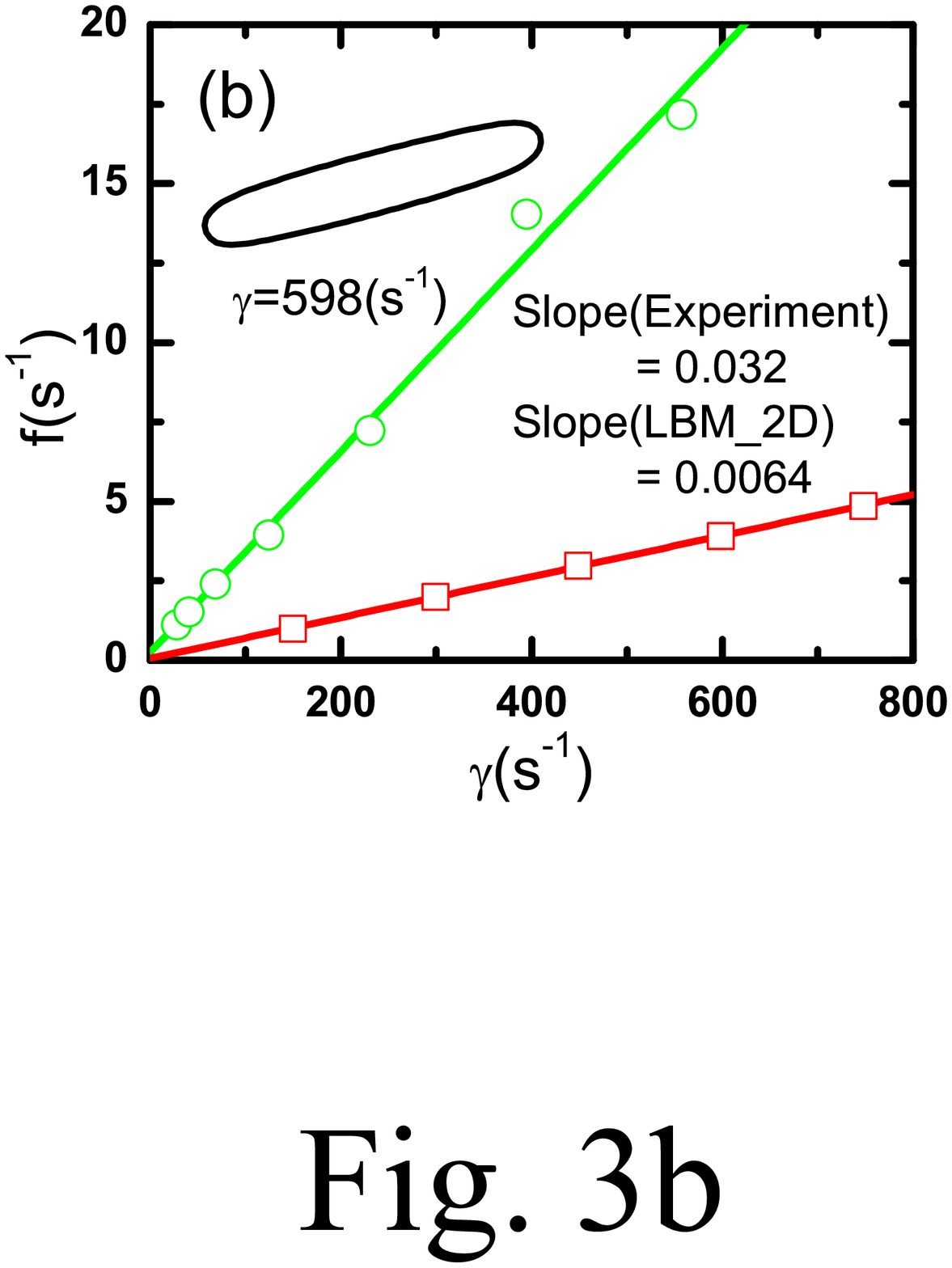}}}
  \end{center}
  \caption{Erythrocyte deforming and moving in shear flow. (a) Orientation angle as a function of the
   swelling ratio $\Gamma$ with the shear rate $\gamma = 7481 \mathrm{s}^{-1}$,together with
    the results of Green function and KS theory~\cite{Beaucourt04}. (b) Tank tread frequency $f$ plotted against the shear rate $\gamma$.
   \textcolor{red}{$\square$} and \textcolor{green}{$\bigcirc$} correspond to lattice
   Boltzmann simulation and experimental results~\cite{Fischer78}, respectively.
   The frequency of lattice Boltzmann simulation is multiplied by $2\pi$.  Inset: a terminal
   snapshot of an erythrocyte in a shear flow for $\gamma = 598$ and
449 $\mathrm{s^{-1}}$.}
  \label{angle}
\end{figure}

%  \caption{Tank tread frequency $f$ plotted against the
 %  shear rate $\gamma$. \textcolor{red}{$\square$} and
 %   \textcolor{green}{$\bigcirc$} correspond to lattice Boltzmann simulation and experimental
 %   results~\cite{Fischer78}, respectively. The frequency of lattice Boltzmann simulation
 %    is multiplied by $2\pi$.  Inset: a terminal
 %   snapshot of an erythrocyte in a shear flow for $\gamma = 598$ and 449 $\mathrm{s^{-1}}$.}
 % \label{rotate}

Now, we performed simulations of an erythrocyte deforming and
moving in a shear flow. According to the KS
theory~\cite{Beaucourt04}, two parameters relevant to the tank
tread-like motion of an erythrocyte are the swelling ratio
\begin{equation}
\Gamma = 4\pi S/P^2,
\end{equation}
and the viscosity ratio
\begin{equation}
r = \rho_{in}\nu_{in}/\rho_{out}\nu_{out},
\end{equation}
where $S$ and $P$ are the area and the girth of a two-dimensional
erythrocyte, $\rho$ and $\nu$ are the density and kinematic
viscosity of the fluid inside or outside the erythrocyte, marked
by subscripts $in$ and $out$, respectively. In our simulations  $r
\approx 1.0$~\cite{PozrikidisBook92}. At first, we pumped out some
fluid inside the erythrocyte until the chemical potential reached
a certain value. Consequently, the upmost and lowest layer in the
simulation domain has a rightward and leftward velocity,
respectively, with a same magnitude to produce a simple shear
flow. Meanwhile, the area of the erythrocyte was kept nearly
invariable by minus increased mass
\begin{equation}
\Delta d=\frac{S\bar{\rho}-S_0\bar{\rho}_0}{M}
\end{equation}
from every fluid node inside the erythrocyte, where $S_0$ and $S$
are the areas of the erythrocyte before and after adding a shear
flow, $\bar{\rho}_0$ and $\bar{\rho}$ are the average densities
inside the erythrocyte before and after adding a shear flow, $M$
is the number of fluid nodes inside the erythrocyte. According to
the KS theory, the orientation angle of an erythrocyte moving in a
shear flow will vary with $\Gamma$. From Fig. \ref{angle} (a), we
can see that our simulation results agree with the results
calculated by Green functions~\cite{Beaucourt04} very well when
the swelling ratio $\Gamma > 0.8$. In fact, when $\Gamma < 0.8$,
the shape of the erythrocyte without any shear flow will be
concave and its terminal shape with a shear flow will be
protuberant, but different from ellipsoid. In Fig. \ref{angle}
(b), the swelling ratio $\Gamma = 0.418$, the erythrocytes undergo
tank tread-like motion, consistent with the famous experimental
observation by Fischer and Schmidt-Sch\"{o}nbein~\cite{Fischer78}.
The final shape of the erythrocyte for $\gamma = 598$
$\mathrm{s^{-1}}$ are shown in the inset of Fig.~\ref{angle} (b).
Numerically we find that the erythrocyte deformed from original
biconcave shape into a highly flattened shape, which the the
boundary integral and singularity methods can not get despite
two-dimensional or three-dimensional
calculation~\cite{Pozrikidis03}. We have computed the tank tread
frequency with respect to the shear rate for $\Gamma = 0.418$. The
result is displayed in Fig.~\ref{angle} (b) together with the
experimental one~\cite{Oka88}. It is clear that our
two-dimensional simulation result has the same linear behavior as
the frequency of experiment. Because the swelling ratio of a cross
section of three dimensional biconcave erythrocytes may
continuously increases to one, our two-dimensional simulation is
only for one certain cross section of three-dimensional, for
example the center cross section.

To summaries, we have developed a lattice Boatsman model to
simulate vesicle deforming and moving in various fluid flows for a
wide range of Reynolds numbers. The results show that the static
biconcave shape is rather steady that it can effectively resist
pathological changes on the membranes. In a shear flow, an
erythrocyte will deform from biconcave shape into a highly
flattened one. Considering the intrinsic parallelity of the
lattice boltzmann method, this scheme may have good prospects for
the simulations of various vesicles deforming and moving in the
vessel or capillary.

This work was partially supported by the National Natural Science
Foundation of China through projects No. 10447001 and 10474109,
Foundation of Ministry of Personnel of China and Shanghai
Supercomputer Center of China.

\end{document}